# Wi-Fi/WiGig Coordination for Optimal WiGig Concurrent Transmissions in Random Access Scenario

[1,2]Ehab Mahmoud Mohamed, [1,3,4]Kei Sakaguchi, and [1]Seiichi Sampei

[1]Graduate School of Engineering, Osaka University, Osaka, Japan. [2]Electrical Engineering Dept., Aswan University, Aswan, Egypt. [3]Tokyo Inastitute of Techology, Japan. [4]Fraunhofer Heinrich-Hertz-Institute, Germany.

Email: ehab@wcs.comm.eng.osaka.ac-u.ac.jp, {sakaguchi, sampei}@comm.eng.osaka-u.ac.jp

*Abstract*— **Wireless Gigabit (WiGig) access points (APs) using 60 GHz unlicensed frequency band are considered as key enablers for future Gbps WLANs. Due to its short range transmission with high susceptibility to path blocking, a multiple number of WiGig APs should be installed to fully cover a typical target environment. However, using autonomously operated WiGig APs with IEEE 802.11ad DCF, the exhaustive search analog beamforming and the maximum received power based autonomous users association prevent the establishment of optimal WiGig concurrent links that maximize the total system throughput in random access scenarios. In this paper, we formulate the problem of WiGig concurrent transmissions in random access scenarios as an optimization problem, then we propose a Wi-Fi/WiGig coordination architecture to solve it. The proposed coordinated Wi-Fi/WiGig WLAN is based on a tight coordination between the 5 GHz (Wi-Fi) and the 60 GHz (WiGig) unlicensed frequency bands. By which, the wide coverage Wi-Fi band controls the establishment of the WiGig concurrent links. Statistical learning using Wi-Fi fingerprinting is used for estimating the best candidate AP and its best beam identification (ID) for establishing the WiGig concurrent link without making any interference to the existing WiGig data links.**

## I. INTRODUCTION

A wireless gigabit (WiGig) indicates the multi-Gbps wireless transmissions in the 60 GHz unlicensed frequency band. IEEE 802.11ad [1] is the currently proposed standard for WiGig transmissions. IEEE 802.11ad defines the use of multi-band (2.4, 5 and 60 GHz) access points (APs) for backward compatibility with legacy IEEE 802.11 a, b, g, and n and to perform fast session transfer (FST) among them. The 60 GHz band suffers from high propagation loss, oxygen absorption and shadowing effect requiring a high directional antenna to overcome these sever channel losses and accomplish data transmission. Accordingly, IEEE 802.11ad defines a MAC based analog beamforming mechanism based on exhaustive searching using switched antenna array to find out the best TX/RX beam directions for a WiGig link and realize user association based on maximum received power.

Due to its small coverage, a numerous number of WiGig APs should be installed in future multi-Gbps WLANs to provide multi-Gbps transmissions for all users inside the WLAN. This can be easily realized using WiGig concurrent transmissions enabled by spatial multiplexing inherent in WiGig directional transmissions [2]. However, establishing optimal concurrent links that maximize the total system throughput in random access scenarios cannot be guaranteed using the currently standardized WiGig APs, i.e., using IEEE 802.11ad with distributed coordination function (DCF). This comes from the autonomous exhaustive search beamforming and the autonomous maximum received power based user association. In random access scenarios using CSMA/CA, the carrier sense (CS) function of a WiGig AP might not indicate the medium busy due to the predominant nature of directional transmissions and receptions. Accordingly, an AP may perform the exhaustive search beamforming while its nearby APs are involved in directional data transmissions. This will cause a lot of packet collisions and interferences to the existing WiGig data links. In addition, due to the randomness inherent in the arrival times of users' packets, WiGig concurrent links are randomly established. As users are autonomously associated based on maximum received power, high mutual interferences occur between these randomly established links. Due to this non-optimality in establishing the WiGig concurrent links in random access scenarios, the total system performance in terms of total system throughput and average packet delay will be highly degraded as we increase the number of autonomously operated WiGig APs.

In this paper, we formulate the problem of WiGig concurrent transmissions in random access scenarios as an optimization problem. In which, we assume that any WiGig AP can establish the concurrent link with the user equipment (UE) in its current transmission opportunity (TXOP) as long as the maximum received power from the AP is sufficient for data transmission. The main target of the optimization problem is to find out the best AP among the candidate un-used APs and the best beam ID (direction) to establish the link with the UE without making any interference to the existing data links while maximizing the total system throughput. To practically solve this optimization problem, in this paper we propose a coordinated Wi-Fi/WiGig WLAN. In which, the wide coverage Wi-Fi (5 GHz) band is used to send control frames required for establishing and managing the low interference WiGig concurrent links. Control frames transmitted by Wi-Fi signaling include frames used to coordinate the beamforming training among the candidate un-used APs. Thus, only one candidate AP can perform the beamforming training at a time. Hence, packet collisions due to simultaneous beamforming training can be effectively eliminated. Also, control frames containing existing links information such as the used APs, the used beam identifications (IDs), the used modulation coding scheme (MCS) indices and the received powers are also broadcasted via Wi-Fi signaling. Therefore, candidate APs can effectively exclude beam IDs (bad beams) that might collide/interfere with the existing data links before conducting beamforming training. This enables the best candidate AP to establish the concurrent link with the UE without causing any interference to the existing data links while maximize the total system throughput as the main target of the optimization problem.

To practically solve the problem of finding out the best candidate APs among the un-used ones and effectively estimate their bad beam IDs, a solution based on statistical learning is proposed in this paper. The proposed approach is based on using Wi-Fi/WiGig fingerprints as that we proposed in [3] [4]. The idea behind this technology is that we can roughly identify the location of a UE by using Wi-Fi channel information measured by multiple APs that is called fingerprint. Since the best candidate APs to establish the link and the best beam ID to be selected are both location dependent. Therefore, finding the best candidate APs and their best communicating beam IDs can be easily estimated if we have a database (DB) to make link between Wi-Fi fingerprints and WiGig best beam IDs. Based on this statistical learning DB, in real-time phase, by just comparing the currently measured UE Wi-Fi fingerprint with the DB, the best candidate APs and a group of best sector IDs (beams) can be estimated for each selected AP to effectively communicate with the UE at its current position. Among these estimated best beams, the beam IDs overlapping with the existing WiGig data links are recognized as bad beams and eliminated from the beamforming training process. These bad beams can be easily estimated using the pre-constructed DB [4].

Via numerical simulations, the proposed coordinated Wi-Fi/WiGig WLAN highly outperforms the IEEE 802.11ad based one and maximizes the performance of WiGig WLANs in random access scenarios.

The rest of this paper is organized as follows; Section II formulates the optimization problem. The proposed coordinated Wi-Fi/WiGig WLAN including the proposed dual band MAC protocol is presented in Sect. III. The performance of the proposed WLAN is analyzed in Sect. IV via numerical simulations. Section V concludes this paper.

## II. OPTIMIZATION PROBLEM FORMULATION

Suppose that there are $M$ APs installed in the WLAN area, and $M_s$ APs are currently used in establishing existing data links. Also, suppose that there is a TXOP for user $k$, and we want to find out the best AP among candidate un-used APs $A(k)$ and its best beam direction that can establish the link with user $k$ without causing any interference to existing links while maximizing total system throughput. $A(k)$ is a subset of total un-used APs domain $\mathbb{C}_{M_{us}}$, i.e., $A(k) \subseteq \mathbb{C}_{M_{us}}$. This optimization problem can be formulated as:

$$\max_{x_{ak(b_a)}} \left( Z_{ak(b_a)} x_{ak(b_a)} + \sum_{\substack{m \neq a \\ n \neq k}} Z_{mn(b_m)} \right), \quad \forall m \in \mathbb{C}_{M_s} \quad (1)$$

s.t.
1. $\sum_{a \in A(k)} \sum_{b_a \in \mathcal{B}_a} x_{ak(b_a)} = 1, \quad A(k) \subseteq \mathbb{C}_{M_{us}}$
2. $MCS_{mn(b_m)}^{x_{ak(b_a)}=1} \not< MCS_{mn(b_m)} \quad \forall m \in \mathbb{C}_{M_s}$
3. $MCS_{x_{ak(b_a)}=1} \geq MCS1$

where $Z_{mn(b_m)}$ indicates the data rate of an existing link between AP $m$ and user $n$ using beam ID $b_m$, and $Z_{ak(b_a)}$ is the data rate given to user $k$ if it is linked to candidate AP $a$ using beam ID $b_a$. $Z_{mn(b_m)}$ can be defined as:

$$Z_{mn(b_m)} = \min(R_{mn(b_m)}, L_n) \quad (2)$$

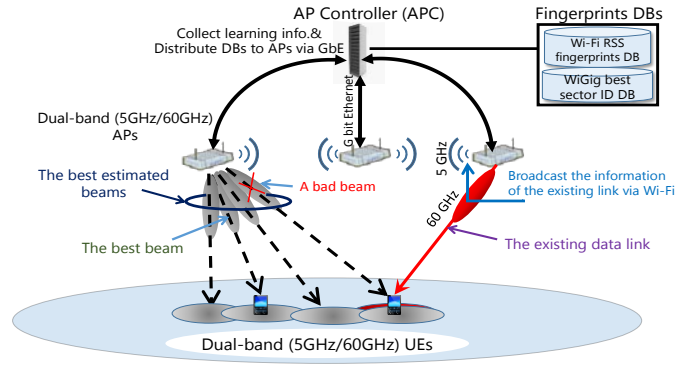

Fig. 1. Coordinated Wi-Fi/WiGig WLAN architecture.

where $R_{mn(b_m)}$ is the achievable transmission rate in bps for user $n$ from AP $m$ using beam ID $b_m$, and $L_n$ is the traffic load of user $n$ in bps. In (2), if the achievable rate is higher than the traffic demand, the user rate is peaked by the traffic demand and vice versa. $R_{mn(b_m)}$ is directly related to the signal-to-interference pulse noise ratio (SINR) from AP $m$ to user $n$ using beam ID $b_m$ [4]. In the optimization equation, $x_{ak(b_a)}$ indicates the link establishment index, which means that if candidate AP $a$ is selected as the best AP and $b_a$ is selected as its best beam ID to establish the concurrent link with user $k$ at its current TXOP, $x_{ak(b_a)} = 1$ and 0 otherwise. The constraint $\sum_{a \in A(k)} \sum_{b_a \in \mathcal{B}_a} x_{ak(b_a)} = 1$ means that only one AP from all candidate APs $A(k)$ and only one beam direction from all candidate beam directions $\mathcal{B}_a$ can be selected to establish the concurrent link with user $k$. The second constraint $MCS_{mn(b_m)}^{x_{ak(b_a)}=1} \not< MCS_{mn(b_m)}, \forall m \in \mathbb{C}_{M_s}$ indicates that the selected link must not collide with any of the existing links by getting its MCS value after establishing the concurrent link $MCS_{mn(b_m)}^{x_{ak(b_a)}=1}$ be less than its value before establishing the link $MCS_{mn(b_m)}$. The final constraint $MCS_{x_{ak(b_a)}=1} \geq MCS1$ means that the MCS value of the selected link should be higher than or equal to the lowest MCS required to accomplish data transmissions which is equal to $MCS1$ [1]. As it is clearly shown, the optimal solution requires an efficient criterion to choose the best candidate APs from the un-used ones that can establish the link with user $k$. Also, it requires a way to organize the process of beamforming training among candidate APs in random access scenarios. Thus, collision free beamforming training among candidate APs can be guaranteed. Also, it requires that the information of current existing data links including the used APs, the used beam IDs and the used MCS values being distributed among installed APs. Thus, candidate APs can successfully eliminate beam IDs that can collide with existing links before starting their beamforming training. Finally, after candidate APs finalize the collision free beamforming training, the best candidate AP and the best beam ID that maximize the total system rate are chosen to establish the concurrent link with the UE. In this paper, we propose a coordinated Wi-Fi/WiGig WLAN to practically fulfill the constraints of the optimization problem and solve it in a distributed manner using random access.

## III. PROPOSED COORDINATED WI-FI/WIGIG WLAN

### A. Proposed WLAN Architecture and Dual Band MAC Protocol

Fig. 1 shows the system architecture of the proposed WLAN using Wi-Fi/WiGig tight coordination. In Fig. 1, multiple dual-band (5 and 60 GHz) APs are connected to an AP controller (APC) via gigabit Ethernet links. The APC works as a central coordinator that assists the establishment of WiGig concurrent links and acts as a gateway to connect the WLAN to the Internet. The wide coverage 5 GHz (Wi-Fi) band in Fig.1 is mainly used to associate the UEs to the WLAN and broadcast the signaling information required for enabling optimal 60 GHz (WiGig) concurrent transmissions with the constraints given in (1). To prevent packet collisions with existing data links as a mandatory constraint of (1), the necessary information of the existing links such as the used APs, the used beam IDs and the used MCS values are broadcasted using Wi-Fi signaling. Therefore, candidate APs can exclude beam IDs (bad beams) that can collide with these existing links before starting beamforming training as shown in Fig.1. In addition, Wi-Fi signaling is used to organize the beamforming operation among the candidate APs. Therefore, the candidate AP which intends to do beamforming training starts the random access process using the Wi-Fi band to prevent any other AP from doing beamforming until it finishes as shown in Fig.1.

To effectively estimate the best candidate APs and estimate their bad beam IDs, statistical learning is used. In the proposed statistical learning approach, the APC collects Wi-Fi received signal strength (RSS) and WiGig best sector ID fingerprints at arbitrary learning points (LPs) in the target environment, and it stores them as fingerprints DBs, see Fig.1. In this paper, we use Wi-Fi RSS as a simple fingerprint of the Wi-Fi signal; other Wi-Fi fingerprinting methodologies can be easily used without any modifications to the general approach [5]. Fingerprints collection is done in an offline phase, which is not repeated unless the transmit power or locations of the APs are changed, or the internal structure of the environment is changed. Then, the APC performs grouping and clustering on the collected Wi-Fi fingerprints to find out the best Wi-Fi fingerprint exemplars for each WiGig best sector ID. Also, the APC finds out the number of offline MCS values provided by each best sector ID via clustering the LPs covered by same best sector ID based on their received MCS values. In the online phase, after collecting and comparing the current UE Wi-Fi RSS readings with the pre-stored Wi-Fi RSS exemplars, the APC estimates groups of best beams from all un-used APs that can communicate with the UE at its current location. The un-used APs with the highest expected MCS values (highest offline MCS values) through their estimated best beams are recognized as the best candidate APs that can establish the link with the UE. Among the estimated best beams, the beam directions that may interfere with the existing links are excluded and recognized as bad beams, as shown in Fig.1. These bad beams are easily estimated using the pre-constructed fingerprint DBs as given in [4], and discussed through the following subsection.

Fig. 2 shows the proposed dual-band MAC protocol, which is used for enabling optimal WiGig concurrent transmissions in random access scenarios using Wi-Fi signaling. In the proposed protocol, if a TXOP is assigned for a specific UE, the APC requests one of the un-used APs to sense the 5 GHz band using CS routine, and if the medium is free, it starts the backoff counter. If the counter reaches zero, it starts to send a Wi-Fi measurement request (Wi-Fi M. Req.) frame to the intended UE. Then, UE broadcasts a Wi-Fi measurement response (Wi-Fi M. Resp.) frame for measuring its current Wi-Fi RSS readings. A switch ON frame is sent to the UE from the un-used AP using its Wi-Fi interface to turn ON its WiGig interface if it is in a sleep mode. Based on its current Wi-Fi RSS readings and offline fingerprint DBs, the APC estimates a groups of best beams to communicate with the UE from all un-used APs. The APs with the highest expected MCS values (highest offline MCS values) are recognized as the best candidate APs. The candidate APs are prioritized by the APC based on their expected MCS values, i.e. the candidate AP having the highest expected MCS value will have the highest priority, etc. The APC also estimates groups of bad beams to make interference with the estimated best beams based on the offline records as explained in [4]. Then, the APC sends this information to the coordinated APs. After this preparation process, the highest priority AP (with the highest expected MCS) after excluding the bad beams starts the random access process in the 5 GHz band. If the backoff counter reaches zero, a NAVset frame is broadcasted by the AP using its Wi-Fi interface to prevent any other candidate AP from doing beamforming training in the 60 GHz band until it finishes. The NAVset frame contains the estimated time that the AP will take until it finishes the beamforming training process. After finalizing beamforming training using beam refinement protocol (BRP) frames and receiving the feedback (FBK) frame from the UE, the first priority AP checks the measured MCS value corresponding to the best communicating beam. If it is sufficient for data transmission, i.e., higher than MCS1, and higher than the expected MCS value of the second priority AP, the first priority AP will establish the link with the UE. In this case, a beam identification (BID) frame containing the actual best beam ID, the actually used MCS index and the actually received power is broadcasted by the AP using the 5 GHz band. By broadcasting the BID frame, other APs will consider the selected link as an existing link and use the broadcasted information when they accurately estimate their bad beams before starting beamforming training. After broadcasting the BID, the AP starts to send data frames to the UE using the estimated best beam in 60 GHz band. A handover is made to the second priority AP to start the beamforming training and try to establish the link if the first priority AP could not find a link with an MCS value sufficient for data transmission and higher than the expected MCS value of the second priority AP. In this case a NACK packet is transmitted from the first priority AP to inform that it fails to establish the link with the UE. Also, a handover is made during data transmission, if the link is blocked. In this case, a NACK packet is transmitted from the UE in the 5 GHz band to start the handover routine using the second priority AP. The process of handover is repeated as long as the NACK packet is sent. If there is no candidate which could establish the WiGig link with the UE, its TXOP will be handled using the Wi-Fi band using FST as proposed by IEEE 802.11ad [1].

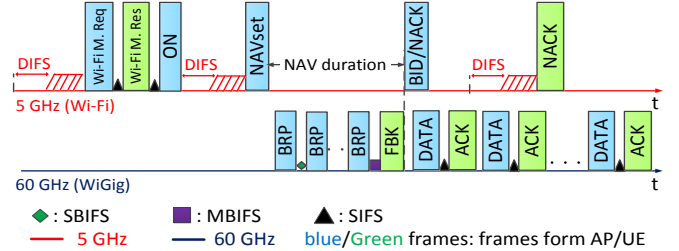

Fig.2. Dual-band MAC protocol for coordinated Wi-Fi/WiGig WLAN.

### B. Statistical Learning for Estimating the Best Candidate APs and Their Best and Bad Beams

In this subsection, we propose to use the statistical learning approach proposed in [3] [4] in estimating the best candidate APs and their best and bad beams. The statistical learning approach consists of two phases; the offline phase and the online phase.

*1) Offline Phase*

The first step in the offline statistical learning phase is to construct the 5 GHz and 60 GHz radio maps for target environments. Constructing the radio maps can be effectively done by collecting the average Wi-Fi RSS readings and the information related to WiGig APs best sectors IDs (fingerprints) at arbitrary LPs in the target environment. Therefore, four databases are constructed as radio maps, i.e., the Wi-Fi fingerprint DB $\mathbf{\Psi}$, the best sector ID DB $\mathbf{\Phi}$, the offline received power DB $\mathbf{P_{OFF}}$, and offline MCS DB $\mathbf{MCS_{OFF}}$ which are defined as:

$$\mathbf{\Psi} = \begin{pmatrix} \psi_{11} & \cdots & \psi_{L1} \\ \vdots & \ddots & \vdots \\ \psi_{1M} & \cdots & \psi_{LM} \end{pmatrix}, \mathbf{\Phi} = \begin{pmatrix} \phi_{11} & \cdots & \phi_{L1} \\ \vdots & \ddots & \vdots \\ \phi_{1M} & \cdots & \phi_{LM} \end{pmatrix},$$

$$\mathbf{P_{OFF}} = \begin{pmatrix} p_{11}^{\phi_{11}} & \cdots & p_{L1}^{\phi_{L1}} \\ \vdots & \ddots & \vdots \\ p_{1M}^{\phi_{1M}} & \cdots & p_{LM}^{\phi_{LM}} \end{pmatrix}, \mathbf{MCS_{OFF}} = \begin{pmatrix} MCS_{11}^{\phi_{11}} & \cdots & MCS_{L1}^{\phi_{L1}} \\ \vdots & \ddots & \vdots \\ MCS_{1M}^{\phi_{1M}} & \cdots & MCS_{LM}^{\phi_{LM}} \end{pmatrix} \quad (3)$$

where $\psi_{lm}$ is the Wi-Fi RSS fingerprint at AP $m$ from a UE located at LP $l$. $L$ is the total number of LPs, and $M$ is the total number of APs. $\phi_{lm}$ is the WiGig best sector ID at LP $l$, which corresponds to the antenna sector ID of AP $m$ which maximizes the received power by a UE located at LP $l$. $\phi_{lm}$ can be calculated as:

$$\phi_{lm} = b_m^* = \arg\max_{b_m}(P_{lm}(b_m)), \ 1 \leq b_m \leq \mathcal{B}_m, \quad (4)$$

where $b_m$ indicates sector ID of AP $m$, $\mathcal{B}_m$ is the total number of sector IDs, and $\phi_{lm} = b_m^*$ is the best sector ID at LP $l$ from AP $m$ that maximizes the received power $P_{lm}(b_m)$. A *null* sector ID in the $\mathbf{\Phi}$ matrix, i.e., $\phi_{lm} = null$, means that AP $m$ cannot cover LP $l$. $p_{lm}^{\phi_{lm}}$ is the power received at LP $l$ from AP $m$ using best sector ID $\phi_{lm}$, and $MCS_{11}^{\phi_{11}}$ is its corresponding MCS index. Clustering is applied on the Wi-Fi fingerprints corresponding to the same best sector ID to reduce the complexity of Wi-Fi fingerprints matching in the online phase. This can be done by finding out the best Wi-Fi fingerprint exemplars $\mathfrak{J}_j^{b_m^*}$, $j \in \{1,2,\ldots,C_{b_m^*}\}$ for each best sector ID $b_m^*$, where $C_{b_m^*}$ is the total number of Wi-Fi fingerprint exemplars (clusters) for $b_m^*$. Then, we find out the offline MCS values achieved by each best sector ID $b_m^*$. This can be done by clustering the LPs covered by the same best sector ID based on their received offline MCS values using $\mathbf{MCS_{OFF}}$ matrix to obtain $MCS_{OFF}^{b_m^*}(s), s \in \{1,2,\ldots,S_{MCS_{OFF}^{b_m^*}}\}$, where $S_{MCS_{OFF}^{b_m^*}}$ is the total number of offline MCS values achieved by best sector ID $b_m^*$. The details of clustering is given in [3].

*2) Online Phase*

During this phase, the online Wi-Fi fingerprint vector $\boldsymbol{\psi}_r$ from the target UE, at an arbitrary position $r$, is measured by the APs and collected by the APC. $\boldsymbol{\psi}_r$ is defined as:

$$\boldsymbol{\psi}_r = [\psi_{r1} \psi_{r2} \ldots \psi_{rM}]^T. \quad (5)$$

Using $\boldsymbol{\psi}_r$, $\mathbf{\Psi}$, $\mathbf{\Phi}$ and $\mathbf{MCS_{OFF}}$, the APC selects the best candidate APs from un-used ones that can establish the link with the UE at its current TXOP. This can be done by first estimating the best beams to communicate with the UE from all un-used APs as follows [4]:

$$b_m^*(1:|\Theta_{b_m^*}|) = \underset{b_m^*}{\text{sort}}\left(\arg\min_{1 \leq j \leq C_{b_m^*}} \left\|\boldsymbol{\psi}_r - \mathfrak{J}_j^{b_m^*}\right\|^2\right)\Big|_{1:|\Theta_{b_m^*}|}, \quad (6)$$

where $\Theta_{b_m^*}$ is the group of best beams for un-used AP $m$ to communicate with the UE at its current position, and (6) is evaluated for $\forall m \in \mathbb{C}_{M_{us}}$. After evaluating $\Theta_{b_m^*}$, the APC selects the best candidate APs $A(k)$ expecting to have the highest MCS values when communicating with the UE. This can be done by selecting the group of APs with the highest offline MCS values using the pre-evaluated $MCS_{OFF}^{b_m^*}$, as follows:

$$A(k)(1:|A(k)|) =$$

$$\underset{m \in \mathbb{C}_{M_{us}}}{\text{sort}}\left(\arg\max_{b_m^* \in \Theta_{b_m^*}}\left(MCS_{OFF}^{b_m^*}\right)\right)\Big|_{1:|A(k)|} \quad (7)$$

After estimating $A(k)$, a priority is given to each candidate AP $a \in A(k)$ by the APC based on its maximum offline MCS value. The APC distributes the information of best candidate APs, their priorities, their best beams, a pre-estimate of bad beam candidates that can collide with them by following the method given in [4] and the offline DBs among the deployed APs. Then, the candidate APs $A(k)$ start to eliminate the bad beams that may interfere with the existing links using the information previously broadcasted by the currently used APs via BID frames. The bad beams elimination is following the constraint given in (1) as follows:

$$b_a^*(i) = \forall_{b_a^*}\left(MCS_{mn(b_m^*)}^{ak(b_a^*)} < MCS_{mn(b_m^*)}\right), \quad (8)$$

$b_a^*(i) \in \Xi_{b_a^*}$, $1 \leq i \leq |\Xi_{b_a^*}|$, $\Xi_{b_a^*} \subseteq \Theta_{b_a^*}$, $a \in A(k), \forall m \in \mathbb{C}_{M_s}$

where $b_a^*(i)$ is a bad beam in the estimated best beams $\Theta_{b_a^*}$ of candidate AP $a$, and $\Xi_{b_a^*}$ is the group of bad beams out of the group of best beams $\Theta_{b_a^*}$ of candidate AP $a$ that might cause an interference to any of the existing links. $MCS_{mn(b_m^*)}$ is the MCS currently used by existing AP $m$ to communicate with UE $n$ using best beam $b_m^*$, and $MCS_{mn(b_m^*)}^{ak(b_a^*)}$ is its expected value if the link $x_{ak(b_a^*)} = 1$ is established and uses best beam $b_a^*$. If the condition $MCS_{mn(b_m^*)}^{ak(b_a^*)} < MCS_{mn(b_m^*)}$ is satisfied, candidate AP $a$ will recognize $b_a^*$ as a bad beam and eliminate it from the beamforming training. The value of $MCS_{mn(b_m^*)}$ is previously broadcasted by AP $m$ when it established the link with UE $n$ via BID frame. To evaluate the condition of $MCS_{mn(b_m^*)}^{ak(b_a^*)} < MCS_{mn(b_m^*)}$, we follow the strict rule given in [4]. In which, candidate AP $a$ tests the condition for all overlapped LPs that are covered by both $b_m^*$ and $b_a^*$ in the $\mathbf{P_{OFF}}$ matrix. If any of the overlapped LPs satisfies this condition, candidate AP $a$ will consider $b_a^*$ as a bad beam. The calculation of $MCS_{mn(b_m^*)}^{ak(b_a^*)}$ is based on calculating the SINR of overlapped LPs using $\mathbf{P_{OFF}}$ matrix as given in [4]. After eliminating the bad beams, candidate APs starts beamforming training based on their priority and the handover decisions to establish the WiGig concurrent link with the UE.

## IV. SIMULATION ANALYSIS

In this section, we test the efficiency of the proposed coordinated Wi-Fi/WiGig WLAN for optimally establishing the WiGig concurrent links and maximizing the total system throughput compared to that using the autonomously operated IEEE 802.11ad APs with DCF.

### A. Simulation Area and Simulation Parameters

Fig. 4 shows a ray tracing simulation area of an indoor office environment. The detailed MAC specifications given in IEEE 802.11ad and IEEE 802.11 standards are simulated, and the steering antenna model defined in IEEE 802.11ad [1] are used as the transmit antenna directivity in the 60 GHz band. Table I gives critical simulation parameters. In the conducted simulations, we concern in measuring the total system throughput [Gbps], which is defined as a sum of throughputs for all successfully delivered packets, and the average packet delay, which is defined as the average time period from the instant when a packet occurs to the instant when the UE completes the reception of it.

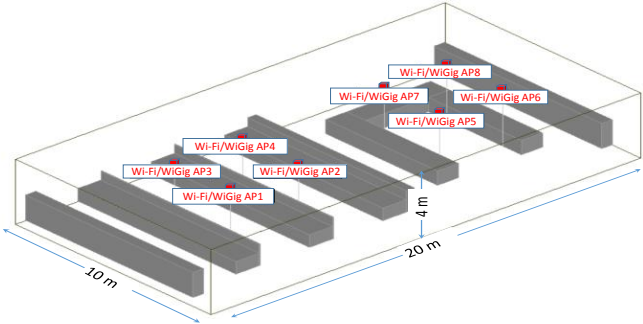

Fig. 3. Ray tracing simulation area.

TABLE I. SIMULATION PARAMETERS

| Parameter | Value |
|---|---|
| Num. of APs / UEs / $|A(k)|$ | 8 / 50 / 2 |
| Tx power of 5 GHz/60 GHz | 20 dBm / 10 dBm |
| Beamwidth in azimuth and elevation directions in 60GHz | 30°, 30° |
| Num. of antenna sectors / beam gain | 36 / 25 dBi |
| Num. of LPs / estimated best beams | 90 / 6 |
| Traffic model | Poisson distribution |
| Offered load per UE / Packet size | Uniform random distribution in the range of [0.5 5] Gbps / 1500 octet |
| Beacon / TXOP interval times | 20 / 1 ms |

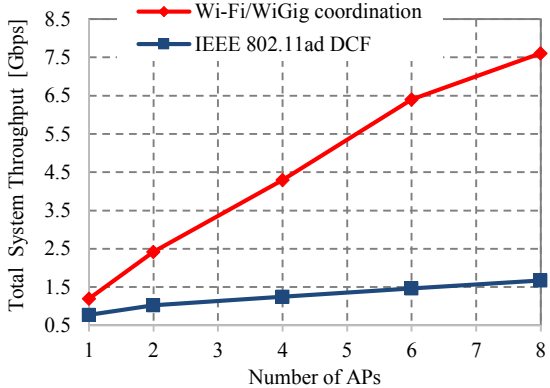

Fig. 4. Average total throughput [Gbps].

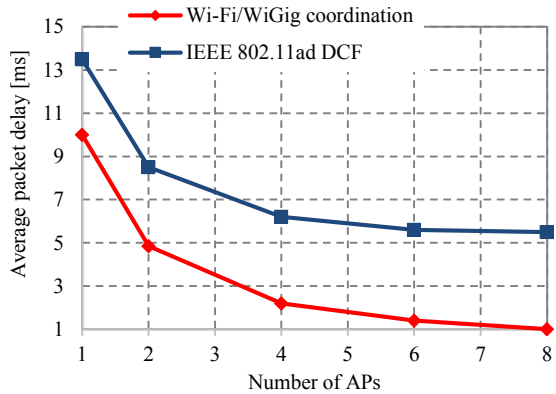

Fig. 5. Average packet delay [ms].

### B. Simulation Results

Figs 4 and 5 show the total system throughput and the average packet delay. In these figures, we compare the performance of the proposed coordinated Wi-Fi/WiGig WLAN with that of the IEEE 802.11ad DCF based WLAN.

In case of single AP, the proposed coordinated Wi-Fi/WiGig WLAN reduces the average packet delay by more than 15% compared to IEEE 802.11ad DCF WLAN. This is because, the statistical learning approach used in the proposed WLAN enables the use of a beamforming mechanism which is faster than the exhaustive search beamforming used in IEEE 802.11ad DCF. This reduction in average packet delay results in increasing the total throughput as shown in Fig. 4.

As we increase the number of used APs, the total throughput of legacy IEEE 802.11ad DCF WLAN is not highly increased even when all 8 APs are operated. This comes from the non-optimality in establishing the WiGig concurrent links due to the un-coordinated exhaustive search beamforming and the maximum power based autonomous users association which cause a lot of packet collisions and interferences in the WLAN. These packet collisions/interferences also affect the average packet delay. On the other hand, using the proposed coordinated Wi-Fi/WiGig WLAN, as we increase the number of used APs, the average total throughput is almost linearly increased. By using 8 APs, about 5-time increase in total system throughput and 6-time decrease in average packet delay are obtained using the proposed coordinated Wi-Fi/WiGig WLAN compared to using IEEE 802.11ad DCF WLAN. This comes from the optimality in establishing the WiGig concurrent links in random access scenarios of which function cannot be provided by IEEE 802.11ad DCF WLAN.

## V. CONCLUSION

In this paper, we proposed an efficient WiGig WLAN based on Wi-Fi/WiGig tight coordination to optimally establish WiGig concurrent links in random access scenarios. The wide coverage Wi-Fi band was used to transmit the signaling information required for establishing and managing the WiGig concurrent links. Statistical learning was proposed to efficiently select the best candidate APs and their best and bad beams. We gave the system architecture of the proposed WLAN in addition to the dual band MAC protocol that organizes the medium access inside the WLAN. Using 8 APs, about 5-time increase in total system throughput and 6-time decrease in average packet delay were obtained using the proposed coordinated Wi-Fi/WiGig WLAN over using IEEE 802.11ad DCF WLAN.

ACKNOWLEDGMENT

This work is partly supported by "The research and development project for expansion of radio spectrum resources" of The Ministry of Internal Affairs and Communications, Japan.

REFERENCES

[1] IEEE 802.11ad standard: "Enhancements for very high throughput in the 60 GHz band," Dec. 2012.
[2] K. Hosoya et al., "Multiple sector ID capture (MIDC): a novel beamforming technique for 60 GHz band multi-Gbps WLAN/PAN systems," *IEEE Trans. Antennas Propagat.*, vol .63, no. 1, pp. 81-96, Jan. 2015.
[3] Ehab M. Mohamed, K. Sakaguchi and S. Sampei, "Millimeter wave beamforming based on WiFi fingerprinting in indoor environment," in *Proc. IEEE ICC Workshops*, Jun. 2015, pp. 1155-1160.
[4] K. Sakaguchi and et al. "Millimeter-wave wireless LAN and its extension toward 5G," *IEICE Trans. on Commun*. Vol E98-B, no.10, Oct. 2010.
[5] C. Feng, W. S. A. Au, S. Valaee, and Z. Tan "Received-signal-strength-based indoor positioning using compressive sensing," *IEEE Trans. Mob. Comput.*, vol. 11, no. 12, pp. 1983-1993, Dec. 2012.